\documentclass[article, 11pt]{amsart}
\pdfoutput=1
\usepackage[english]{babel}
\usepackage[utf8x]{inputenc}
\usepackage[T1]{fontenc}
\usepackage{float}
\usepackage[top=2cm,bottom=2cm,left=3cm,right=3cm,marginparwidth=2cm]{geometry}
\usepackage{amsmath}
\usepackage{color}
\usepackage[version=4]{mhchem} 
\usepackage[colorinlistoftodos]{todonotes}
\usepackage[colorlinks=true, allcolors=black]{hyperref}
\usepackage{xcolor}
\usepackage{siunitx}

\begin{document}
\begin{center}

\title{Spin-Acoustic Control of Silicon Vacancies in 4H Silicon Carbide}
\maketitle

% Author names
\large
Jonathan R. Dietz$^{\dagger1}$, Boyang Jiang$^{\dagger2}$, Aaron M. Day $^{1}$, Sunil A. Bhave$^{2}$ and Evelyn L. Hu$^{1}$ \\

% Author affiliations and notes
\small
$^1$ John A. Paulson School of Engineering and Applied Sciences, Harvard University, Cambridge, MA. 02138, USA \\
$^2$ OxideMEMS Lab, Purdue University, West Lafayette, IN. 47907, USA \\
$^{\dagger}$ Equally contributing authors \\

\end{center}

\begin{abstract}
    We demonstrate direct, acoustically mediated spin control of naturally occurring negatively charged silicon monovacancies (V$_{Si}^-$) in a high quality factor Lateral Overtone Bulk Acoustic Resonator fabricated out of high purity semi-insulating  4H-Silicon Carbide. We compare the frequency response of silicon monovacancies to a radio-frequency magnetic drive via optically-detected magnetic resonance and the resonator's own radio-frequency acoustic drive via optically-detected spin acoustic resonance and observe a narrowing of the spin transition to nearly the linewidth of the driving acoustic resonance. We show that acoustic driving can be used at room temperature to induce coherent population oscillations. Spin acoustic resonance is then leveraged to perform stress metrology of the lateral overtone bulk acoustic resonator, showing for the first time the stress distribution inside a bulk acoustic wave resonator. Our work can be applied to the characterization of high quality-factor micro-electro-mechanical systems and has the potential to be extended to a mechanically addressable quantum memory.  
\end{abstract}

\section{Introduction}

Acoustic control of semiconductor defect spins is by now well established as an attractive complement to magnetic control in sensing and quantum information processing (QIP) \cite{maity2020coherent}. Acoustic driving can bridge the full ground state spin manifold of several defect systems with spin greater than spin-1/2 \cite{ whiteley_spinphonon_2019, macquarrie_coherent_2015, hernandez-minguez_anisotropic_2020}. Hybrid systems that employ acoustic and magnetic control of defect spin states provide a means to transduce spin signals and facilitate integration of spin systems with more conventional technology \cite{lee_topical_2017}. Silicon carbide is an ideal material platform for developing such a hybrid quantum system as it already is capable of high performance optical, acoustic, and electronic devices, having the potential to be an ‘all-in-one’ wafer-scale QIP platform \cite{lukin_4h-silicon-carbide--insulator_2020, bracher_selective_2017, wolfowicz_electrometry_2018,Wang2021}. In addition, acoustic control of defect spins is also useful for the metrology of microelectromechanical system (MEMS) devices in materials like silicon carbide, which hosts several stress-sensitive spin active defects \cite{whiteley_spinphonon_2019,soykal_quantum_2017, udvarhelyi_ab_2018}. Indeed, diamond based sensor have already been used in a variety of applications \cite{schirhagl_nitrogen-vacancy_2014}. Although a variety of techniques can be used to analyze the performance of RF MEMS resonators and filters, these have limited capability to probe the actual distribution of strain within the resonator material \cite{hengky-heterodyne-2009}. Such knowledge is critical in optimizing resonator design and in improving quality factor and power handling. This work demonstrates acoustically mediated spin control of naturally occurring silicon monovacancies (V$_{Si}^-$) in a high-quality factor Lateral Overtone Bulk Acoustic Resonator (LOBAR) fabricated in high purity semi-insulating (HPSI) 4H-Silicon Carbide (SiC). These experiments provide a first step towards high resolution-spatial mapping of strain inside a MEMS resonator, and lay the foundation for the direct integration of spins into a wafer scale material platform for quantum information processing.

Silicon monovacancies are a long-studied defect system in silicon carbide as a spin-photon interface \cite{riedel_resonant_2012,nagy_quantum_2018,nagy_high-fidelity_2019}. The singly charged k-site silicon monovacancy (V$_{Si}^-$) is particularly attractive due to its bright, near-infrared optical transition, often labelled V2 in literature, and its spin-3/2 ground state manifold that can be manipulated through its room-temperature optically-detected spin-acoustic resonance \cite{hernandez-minguez_anisotropic_2020}. We seek to monitor the V$_{Si}^-$s within acoustic resonators where most of the mechanical strain energy is stored in single-crystal material (4H-SiC, in this case), rather than in the piezoelectric transducer. SAW cavity resonators satisfy this criterion, and a Gaussian SAW resonator has been used to drive spin resonances in silicon vacancy and divacancy spin ensembles in 4H-SiC \cite{whiteley_spinphonon_2019, hernandez-minguez_acoustically_2021}.  However, in this case the mechanical strain energy is concentrated near the resonator surface, making the SAW cavities’ quality factors (Q) and frequency susceptible to changes in surface conditions \cite{flannery_effects_2002}. In addition, the coherence of color centers themselves is also sensitive to surface conditions and the presence of stray electric fields, making it desirable to develop devices that make emitters locally addressable in bulk material, a goal which bulk acoustic resonators can enable \cite{rosskopf_investigation_2014,chen_acoustically_2020}.

Therefore, we have chosen to utilize LOBAR, which preserve strain in the entire resonator body, rather than on the surface alone as a means of demonstrating this bulk sensitivity. LOBARs are overtone acoustic wave resonators, where the acoustic cavity length (and hence the resonances) is controlled by the lateral, lithographically-defined extent of the resonator. SiC LOBARs have been demonstrated with mechanical quality-factors of 3000 to 100,000 with frequencies ranging from MHz to GHz \cite{gong_175_2011,ziaei-moayyed_silicon_2011,jiang_semi-insulating_2021}. In this work we study the spatial characteristic of the acoustic coupling of LOBARs to an ensemble of silicon monovacancies, demonstrate spatial and directional strain sensitivity and extend this to show the first example of coherent acoustic control of silicon monovacancies in silicon carbide. 

\begin{figure}[h]
    \centering
    \includegraphics[scale = 0.25]{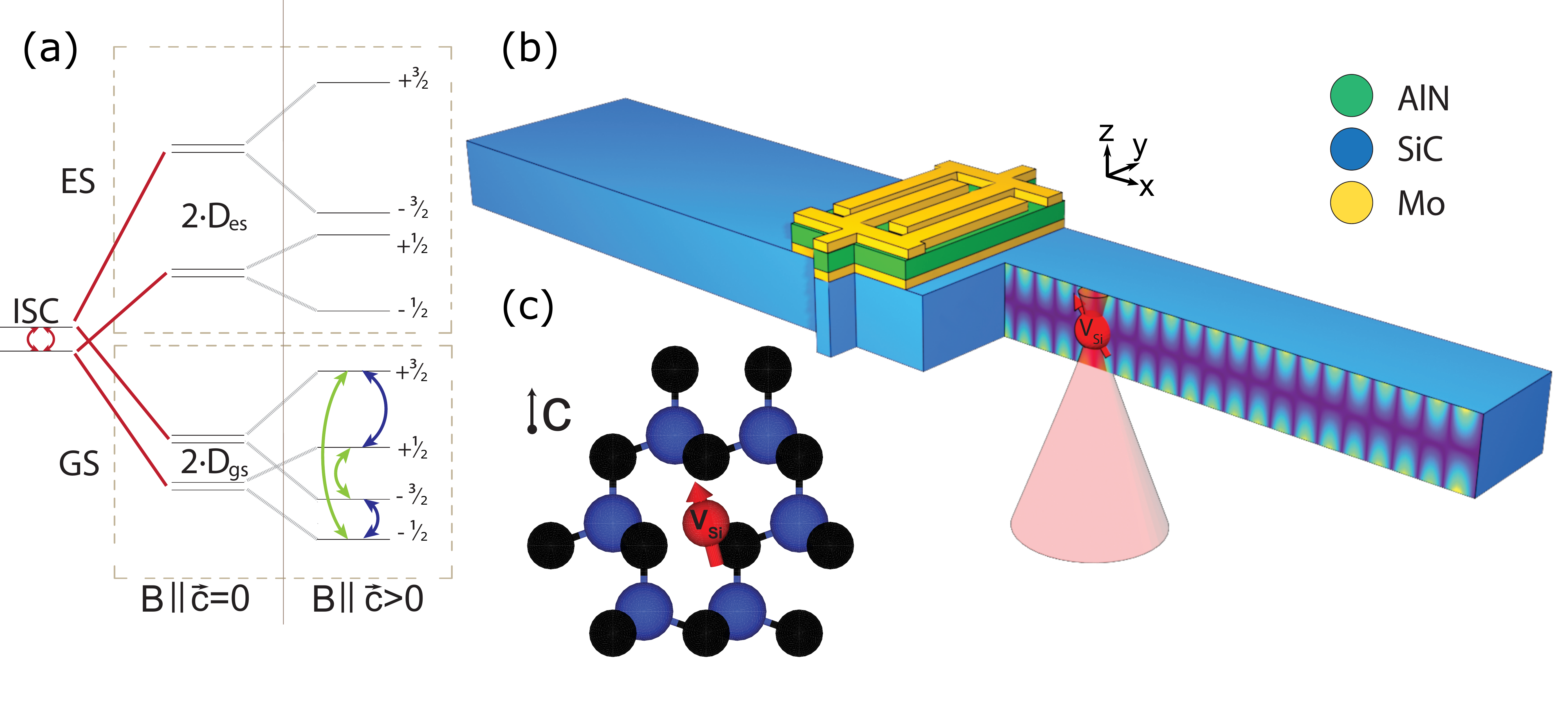}
        \caption{
        (a.) Electronic structure of V$_{Si}^-$ with magnetically-allowed and magnetically-forbidden but acoustically-allowed transitions labelled in blue and green respectively. (b.) A cut-away schematic of the Lateral Overtone Bulk Acoustic Resonator device showing the transducer geometry, back-side illumination, and an example strain distribution from finite element analysis. (c.) Crystallography of the k-site V$_{Si}^-$ (V2) defect center, with nearest neighbor carbon and silicon atoms shown in black and blue respectively.
        }
    \label{F1}
\end{figure}
\color{black}

\section{Results}
V$_{Si}^-$ is present as a naturally-occurring, intrinsic point defect in our LOBAR devices, giving us the ability to characterize the coupling between its bulk acoustic mode and the V$_{Si}^-$ ground state spin (Fig. \ref{F1}a). To unequivocally demonstrate spin-acoustic coupling in our devices, we apply strain to the ensemble alternately with a differentially driven acoustic resonance via  interdigital electrodes (IDE) at its 89.5MHz flexural mode or with magnetic driving via a nearby suspended radio frequency (RF) antenna (Fig. \ref{F1}b). Spin-3/2 defects have a spin-stress coupling Hamiltonian: \cite{soykal_quantum_2017,udvarhelyi_ab_2018,poshakinskiy_optically_2019}

\begin{equation}
H_\epsilon = \Xi \sum_{ij} \epsilon_{ij} S_i S_j
\end{equation}

where $\Xi$ is the interaction strength, $\epsilon_{ij}$ is the stress coupling tensor and $S$ are the spin-3/2 spin matrices. We perform our measurements by tuning the frequency of the ground state spin resonances with a magnetic field oriented parallel to the z-axis of the resonator and c-axis of the SiC (Fig. \ref{F1}b). In this orientation $\Delta m_{s}$=±1 is driven by $(\epsilon_{xx}-\epsilon_{yy})$, $\epsilon_{xy}$, $\epsilon_{xz}$ and $\epsilon_{yz}$ terms, and $\Delta m_{s}$=±2 is entirely driven by $\epsilon_{xy}$ and $\epsilon_{yz}$ terms.\cite{soykal_quantum_2017} We then collect the differential fluorescence of the k-site V$_{Si}^-$ (Fig. \ref{F1}c), from 900-1150nm to measure optically detected spin acoustic resonance (ODSAR). We focus on the asymmetric flexural mode propagating through the LOBAR, since it has higher piezoelectric coupling coefficient, $k_t^2$, and thus greater energy transferred from the piezoelectric transducer to the SiC \cite{jo_d15-enhanced_2017}.

\begin{figure}[h]
    \centering
    \includegraphics[scale=0.8]{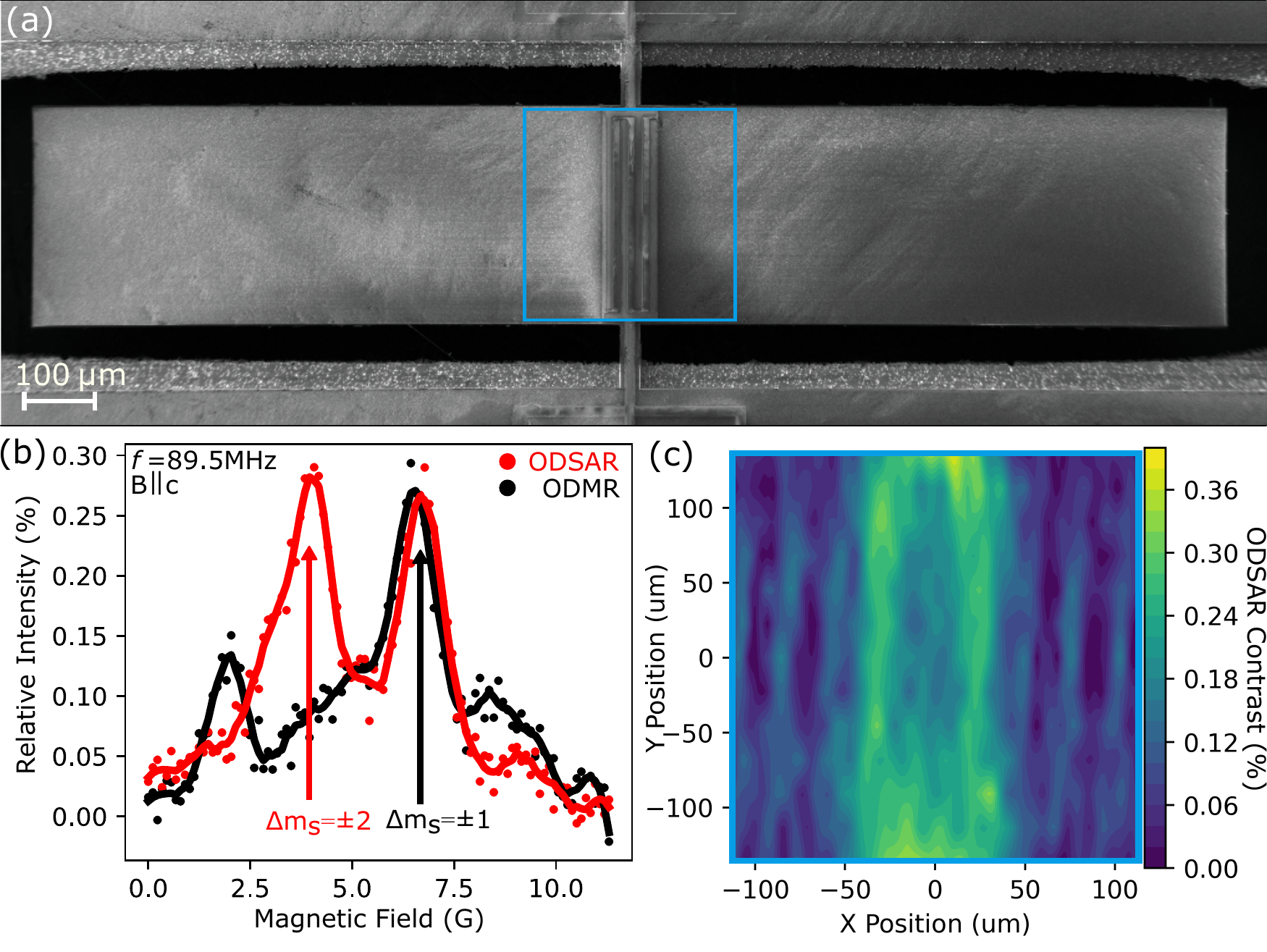}
        \caption{(a.) Field Emission Scanning Electron Microscope image of the actual LOBAR device measured in this experiment (b.) Comparison of spin resonance driven at 89.5MHz with a radio-frequency antenna (Optically Detected Magnetic Resonance) against the LOBAR's transducer (Optically Detected Spin-Acoustic Resonance) showing clearly distinguished peaks associated in energy with spin-1 magnetically allowed transition and a spin-1 and spin-2 acoustically allowed transitions. (c.) XY raster over the area marked in (a) showing fringes characteristic of the flexural modes of the LOBAR device with the $\Delta m_s=2$ transition. }
    \label{F2}
\end{figure}

In our analysis, we differentiate between a regions of the device shown in Fig. \ref{F2}a. These are the \emph{transducer}, where the Mo-AlN-Mo piezotransducer covers the SiC resonator, the \emph{wings}, where the SiC is free of the transducer, and the \emph{tethers}, where the bar is connected to the substrate. To improve signal to noise, a large pinhole of 50um is employed which ensures that the confocal slice of the microscope averages over the full 10um device height during the measurements. From the ODSAR and optically detected magnetic resonance (ODMR) underneath the transducer in Fig. \ref{F2}b, two peaks appear in the ODSAR spectrum which correspond to $\Delta m_{s}$=±2, which are not magnetically dipole-allowed for the corresponding ODMR spectrum. 2D scans of the transducer and nearby wings reveal a characteristic fringe pattern when transduction is on resonance with one of the resonator's overtones Fig. \ref{F2}c. We also note that magnetic driving of the spin transition by stray RF fields from the IDE is suppressed because the SiC resonator is shielded from RF radiation by the presence of a ground plane underneath the signal electrodes having zero piezoelectric displacement current.

From \ref{F2}.c. it is clear that device stress is highly concentrated in the direct vicinity of the transducer, with a slightly lower ODSAR contrast observed on the wings of the device. The flexural modes in the resonator address a smaller sub-ensemble of the defects probed by our confocal slice, resulting in lower average stress over the depth slice. The higher stress under the transducer indicates that a larger ensemble is addressed by the acoustic waves, showing that spin dynamics underneath the transducer are driven by broadband piezo-electric actuation, while the dynamics on the wings are driven by acoustic standing waves that span the entire device. The increase in ODSAR intensity near the tethers of the device quantitatively confirms that stress is concentrated between the suspended resonator and the surrounding substrate, providing a quantitative spatial measure of acoustic loss in the LOBAR.

\begin{figure}[h]
    \centering
    \includegraphics[scale = 0.8]{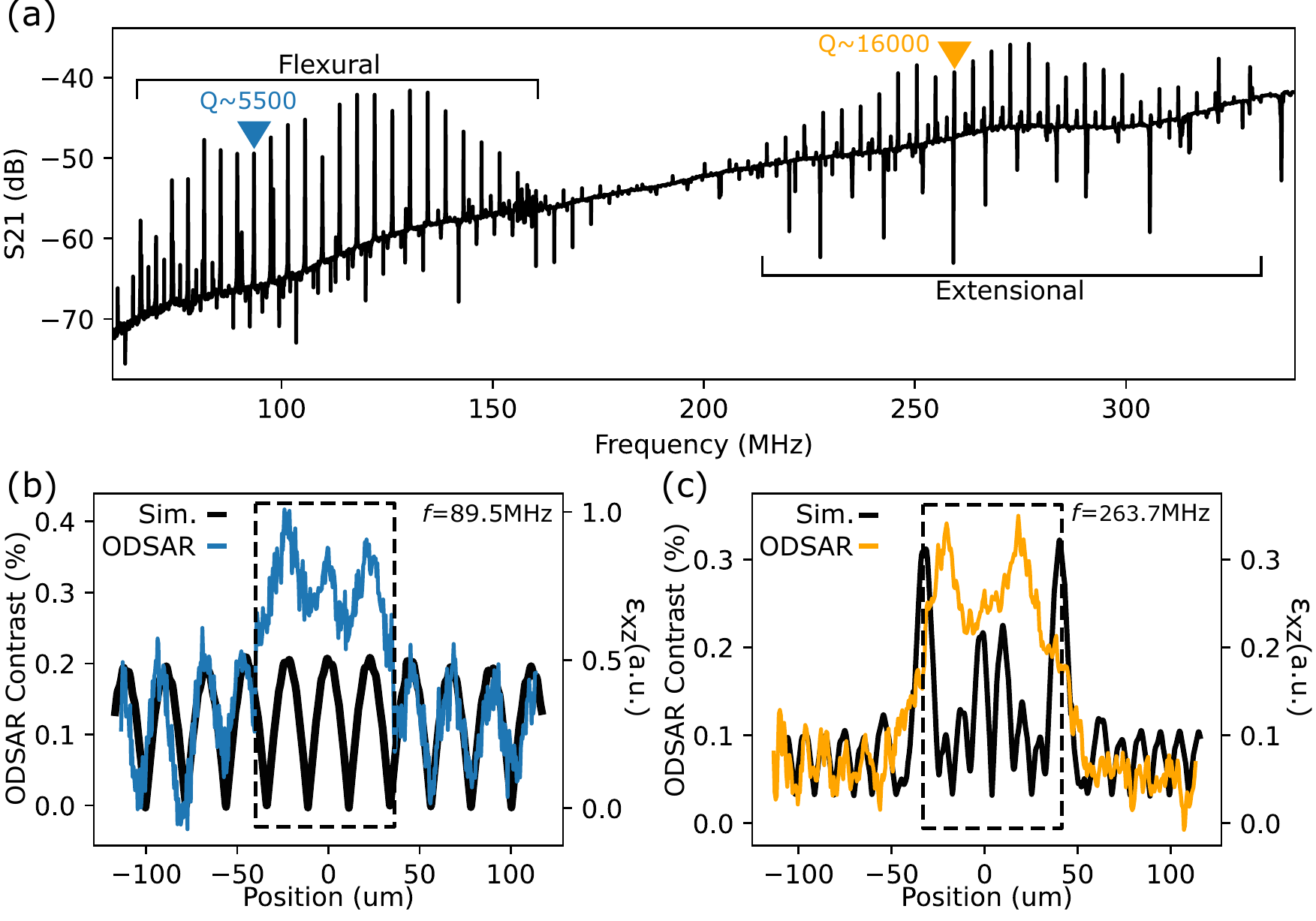}
        \caption{(a.) The $S_{21}$ electrical properties of the resonator in air at room temperature, showing two overtone regions of flexural and extensional acoustic mode families, with the modes shown in (b) and (c) marked with their mechanical Q. Device line scan along the x-axis of the device (Fig.1b) showing ODSAR intensity across the transducer and nearby wing regions of the device, with comparison to simulated predictions for the shearing stress $\sigma_{xz}$ (b.) 89.5MHz flexural and  (c.) 260.7MHz extensional modes of the device. Underneath the transducer (dotted black box) ODSAR deviates from simulation as a result of off-resonant piezo-transduction.
        }
    \label{F3}
\end{figure}

Next, we demonstrate that our technique is capable of spatially resolved metrology of stress in the bulk of the LOBAR by performing spatial confocal scans of the spin resonance intensity of mechanical modes. From the electrical data in Fig. \ref{F3}a we observe the presence of two mode families corresponding to low-frequency flexural ($A_0$) and high-frequency extensional ($S_0$) modes, whose presence are predicted by finite-element model (FEM) simulations. Flexural modes have out-of-phase bulk stress modes that result in a high overall shear stress. By contrast, extensional modes have high in-phase bulk stress but minimal shear stress, excepting the shear stress that is introduced by symmetry breaking introduced by the transducer. In the subsequent work we focus on $\Delta m_{s}$=±1 transitions because they couple both the shear and bulk stress and allow a comparison to the achievable Rabi rates with a standard magnetic drive.

ODSAR can also be used to study the depth-averaged device stress for different resonant mode families. We first identify two high quality factor modes corresponding to the flexural and extensional mode families of the device (Fig. \ref{F3}a). 1D scans along the length of the resonator demonstrate a clearly differentiated spatial frequency and driving dynamics for the 89.5MHz flexural and 263.7MHz extensional modes of the resonator (Fig. \ref{F3}b and \ref{F3}c). With $ \vec{B} \parallel \vec{c}$, the tuned $\Delta m_s=1$ demonstrates a significantly higher spin strain coupling for high-shear stress flexural modes. In the case of the flexural mode, this is confirmed, as the ODSAR shows excellent agreement with the shear strain of the resonator predicted by FEM simulation (Fig. \ref{F3}b). By contrast, the extensional mode of the resonator is expected to have minimal shear stress, except for shear stress that is introduced by the symmetry breaking effect of the transducer, which our scan confirms (Fig. \ref{F3}c). We can also use our technique to probe the local acoustic dynamics introduced by etching asperities on the device \cite{supplement}. Local asperities cause stress localization, which leads to mechanical dissipation or a drop in quality factor. Such a spatially resolved stress localization is critical to identifying "quality factor loss" regions and will provide feedback for improving the MEMS manufacturing technology.
\begin{figure}[h]
    \centering
    \includegraphics[scale = 0.8]{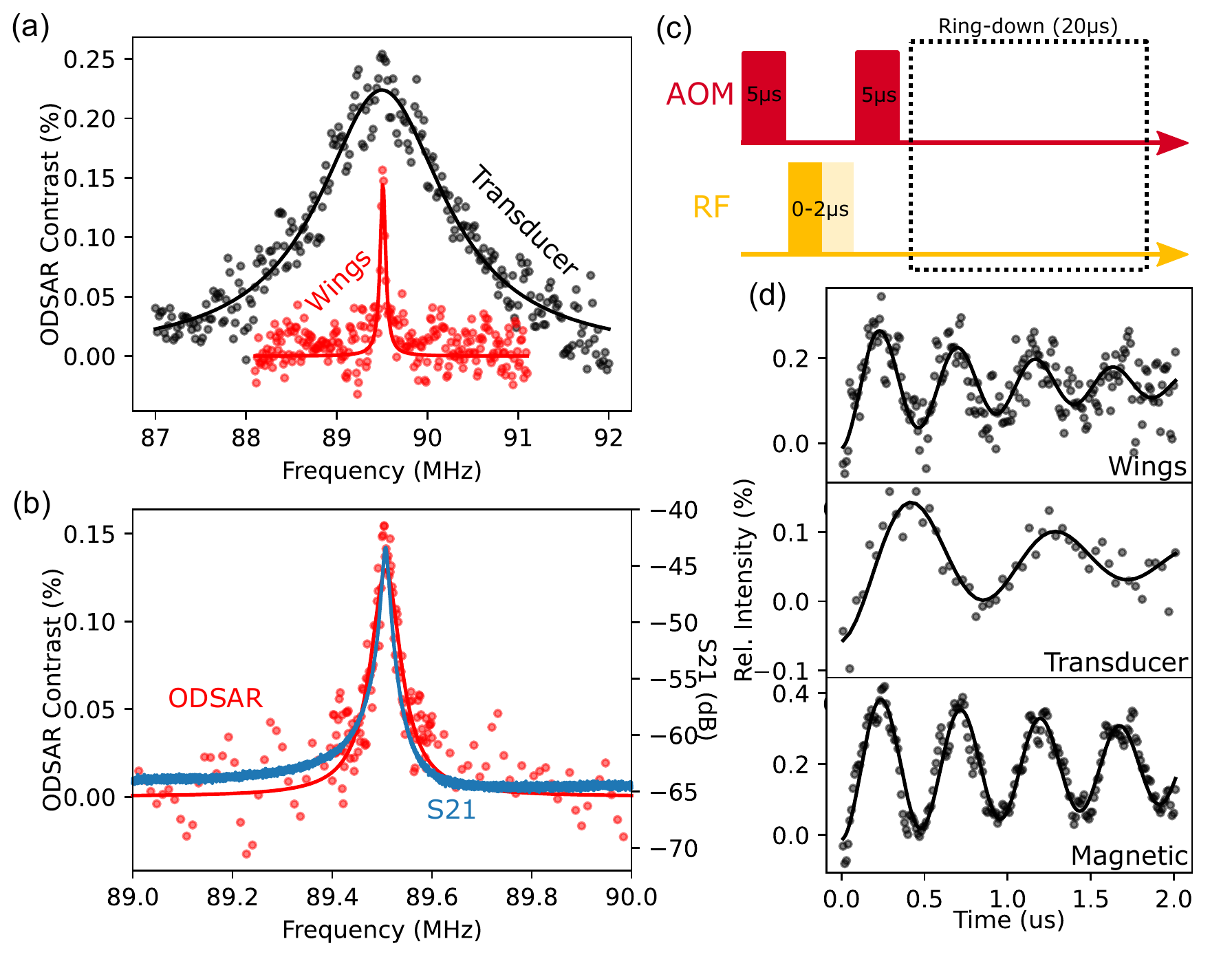}
        \caption{(a.) Comparison of ODSAR signal on the area of the device directly underneath the transducer (“Transducer”) and on the bulk of the resonator separated from the transducer (“Wings”). (b.) Comparison of high resolution electrical measurement of quality factor and measurement of quality factor via the spin-acoustic resonance, showing effective filtering of the spin resonance by the acoustic cavity. (c.) Pulse sequence used to measure coherent population oscillations followed by ring-down of the acoustic cavity shown. (d.) Rabi oscillations measured using an acoustically resonant RF drive (89.5MHz) on the bulk of the resonator (“Wings”), using an acoustically off-resonant but spin-resonant RF drive (89.55MHz) directly underneath the transducer ("Transducer"), and spin-magnetic drive (89.5MHz) from a closely placed RF antenna ("Magnetic"). }
    \label{F4}
\end{figure}

Finally, we demonstrate that ODSAR can be exploited for room temperature coherent control, demonstrating that ground state acoustic driving could be useful for state preparation and as a resource for entanglement generation \cite{PhysRevX.12.011056}. We first characterize the difference in the frequency domain line shape of the ODSAR in Fig. \ref{F4}a. On the transducer, the ODSAR line is a broad 2MHz resonance, while on the wings of the device, the line is narrowed significantly, approaching the linewidth of the mechanical resonance (Fig. \ref{F4}b). Based on this insight, we compare Rabi oscillations directly underneath the transducer and on the wings with the same RF drive power, by initializing the spins with an initial laser pulse, followed by RF excitation at acoustic resonance with a varying pulse length (Fig. \ref{F4}c). Under the wings, spins oscillate at 2.0±0.2MHz, compared to spins under the transducer which oscillate from off-resonant piezo-transduction at 1.2±0.2MHz. Stronger Rabi oscillations on the wings of the device indicate that while the depth-integrated stress experienced by the ensemble is lower, at stress maxima the stress intensity is higher. We attribute the faster decay in the oscillation when compared to magnetically driven Rabi oscillation to the inhomogeneity in the acoustic stress distribution through the depth of our confocal slice.

\section{Discussion}
We have shown that bulk acoustic modes and V$_{Si}^-$ can be acoustically coupled via spin acoustic resonance, and have exploited this resonance to study the dynamical acoustic performance of a LOBAR. Our study finds that spins can be driven both by off-resonant forced oscillations and acoustic standing-wave resonance. An increased Rabi oscillation rate and filtered spin response highlights the benefit of using resonant elements in transduction schemes. Current studies are limited to ensemble measurements as a result of the naturally occurring V$_{Si}^-$ in the HPSI 4H-SiC used. Higher depth resolution can be achieved through layered ion implantation into material with fewer intrinsic defects in the unimplanted material, useful for characterizing acoustic systems where stress is not concentrated near the surface of the device \cite{kasper-irradiation}. As a MEMS metrology technique, this could see application in monitoring the performance of accelerometers, gyroscopes, and clocks over their lifetime and allows a fully depth sensitive means of device characterization \cite{zhai-design-2022,hamelin-monocrystalline-2021,gosavi-hbar-2015}. The identification of device tethers as a center of stress concentration could contribute to improvements in quality factor via tether optimization \cite{hamelin2019monocrystalline}. Conversely, high purity or isotopically purified material with a minimal ensemble density would also open studies of single defect acoustic coupling as a demonstration for quantum information processing applications. Thinning devices to smaller thickness could help increase acoustic coupling and enhance the rate of optical read-out significantly \cite{ ziaei-moayyed_silicon_2011,dietz2022optical}. This experimental work also reinforces recent theoretical proposals suggesting V$_{Si}^-$ as a potential candidate for hybrid quantum memories and a means of microwave to optical qubit transduction schemes in a future quantum network \cite{raniwala-2022-spin}.

\section{Methods}
The LOBAR devices are made on a \SI{500}{\micro\metre} thick semi-insulating 4H-SiC wafer from Cree Inc. Then, it is thinned down to \SI{200}{\micro\meter} using chemical mechanical polishing (CMP), before it is sent to Plasma-Therm for AlN deposition. A Mo-AlN-Mo layer of \SI{100}{\nano\meter}:\SI{1000}{\nano\meter}:\SI{100}{\nano\meter} thickness is deposited with sputtering. The LOBAR is fabricated with a combination of dry and wet etching, with further detail found in \cite{jiang_semi-insulating_2021}. Finite Element Simulations are made using the COMSOL Multiphysics package to simulate device performance. Spin measurements are performed at room-temperature with a home-built confocal microscope \cite{supplement}. The ensemble is excited by a 2mW 865nm external cavity diode laser focused by a 0.9NA microscope objective onto the backside of the LOBAR device. Ensemble photoluminescence is collected by the same objective and filtered by a \SI{900}{\nano\meter} dichroic, a \SI{900}{\nano\meter} long-pass filter, and a \SI{1000}{\nano\meter} short-pass filter before being focused onto a \SI{50}{\micro\meter} pinhole and then collected by a single-photon avalanche-diode detector. Free space magnetic radio-frequency excitation is supplied by \SI{25}{\micro\meter} wire suspended nearby and acoustic radio-frequency excitation is applied through wire-bonded ground-signal-ground pads in a two port, differential driven configuration. For ODSAR and ODMR, the input microwave power is 22dBm, for Rabi oscillation measurements, the input power is 32dBm.
\section{Acknowledgements}
The research was supported by NSF RAISE-TAQS Award 1839164. A.M.D. acknowledges support from the Science and Technology Center for Integrated Quantum Materials, National Science Foundation (NSF) Grant DMR-1231319. The SiC resonators were fabricated in the Birck Nanotechnology Center at Purdue University. The CMP was completed at NOVASiC and the AlN deposition was performed by PlasmaTherm. We thank Dr. Oney Soykal for helpful discussions.

\bibliographystyle{ieeetr}
\bibliography{PurdueLib}
\end{document}